\documentclass[aps,prb,twocolumn,amsmath,amssymb,superscriptaddress,floatfix]{revtex4}
\usepackage{graphicx}
\usepackage{bm}
\usepackage{amssymb}
\usepackage{color}
\usepackage{amsmath}
\newcommand{\be}{\begin{equation}}
\newcommand{\ee}{\end{equation}}
\newcommand{\beqn}{\begin{eqnarray}}
\newcommand{\eeqn}{\end{eqnarray}}

\begin{document}

\title{Boundary critical phenomena of the random transverse Ising model in $D \ge 2$ dimensions}

\author{Istv\'an A. Kov\'acs}
\email{kovacs.istvan@wigner.mta.hu}
\affiliation{Wigner Research Centre, Institute for Solid State Physics and Optics,
H-1525 Budapest, P.O.Box 49, Hungary}
\author{Ferenc Igl\'oi}
\email{igloi.ferenc@wigner.mta.hu}
\affiliation{Wigner Research Centre, Institute for Solid State Physics and Optics,
H-1525 Budapest, P.O.Box 49, Hungary}
\affiliation{Institute of Theoretical Physics,
Szeged University, H-6720 Szeged, Hungary}

\date{\today}

\begin{abstract}
Using the strong disorder renormalization group method we study numerically the critical behavior of
the random transverse Ising model at a free surface, at a corner and at an edge in
$D=2,~3$ and $4$-dimensional lattices. The surface magnetization exponents are found to be:
$x_s=1.60(2),~2.65(15)$ and $3.7(1)$ in $D=2,~3$ and $4$, respectively, which do not depend on the
form of disorder. We have also studied critical magnetization profiles in slab, pyramid and wedge
geometries with fixed-free boundary conditions and analyzed their scaling behavior.
\end{abstract}

\maketitle

\section{Introduction}
\label{sec:intr}
The quantum Ising model with random couplings and/or with random transverse fields (RTIM) is the
prototype of disordered quantum magnets having discrete symmetry. This model has a zero-temperature
quantum phase transition, the properties of which have been studied by a special strong disorder
renormalization group (SDRG) method\cite{im}. In this method the strongest local terms of the Hamiltonian are
successively eliminated and at the same time new terms are generated perturbatively between remaining degrees of
freedom\cite{mdh}. In one dimension (1D) where the topology of the lattice stays invariant under the
transformation the SDRG equations have been solved analytically in the vicinity of the quantum
critical point\cite{fisher}. In this case the phase transition is shown to be controlled by a so called
infinite disorder fixed point\cite{danielreview} (IDFP), in which disorder fluctuations are completely
dominant over quantum fluctuations and therefore the renormalization steps are asymptotically exact for
large scales. Indeed the SDRG results in 1D are consistent with findings of other
analytical\cite{mccoywu,shankar} and numerical methods\cite{young_rieger96,igloi_rieger97,bigpaper}.

In higher dimensional lattices the topology of the lattice is changed during the SDRG steps, therefore
the SDRG method has to be implemented numerically. The first numerical calculations have been performed in
2D\cite{motrunich00,lin00,karevski01,lin07,yu07,ladder}
and more recently an efficient numerical algorithm\cite{2dRG,ddRG,ddRG2} of the present authors made possible to extend
the calculations\cite{ddRG,ddRG2,phd,kovacs_igloi12} to 3D and 4D, as well as to Erd\H os-R\'enyi random graphs,
which are infinite dimensional
lattices. In all dimensions the phase transition is found to be controlled by an IDFP, which
justifies that the SDRG method provides asymptotically exact results for large systems in higher
dimensions, too. Quantum Monte Carlo simulations for the 2D RTIM are consistent with the SDRG results\cite{pich98}. Similarly,
simulation results for the random contact process\cite{vojta09,vojta12} - which is expected to be in the same universality class\cite{hiv}
as the RTIM - are in agreement with the SDRG results in 2D and in 3D.

In $D>1$ dimensions during the SDRG iterations a large number of new couplings
are generated between remote sites, which makes the numerical implementation of the method
rather cumbersome. To avoid this problem another, more simple approximation methods have been developed and applied to the
RTIM\cite{dimitrova_mezard,monthus_garel1,monthus_surface,monthus_garel2,monthus_garel3,monthus_garel4,miyazaki_nishimori}.
One of those\cite{dimitrova_mezard} is based on the quantum cavity approach\cite{cavity},
which is found to reproduce some of the exact results in 1D.
However, in the Bethe
lattice with an effective dimensionality of $D_{\text{eff}}=2$ the method has predicted conventional random critical behavior
instead of IDFP scaling. The quantum cavity method is shown to be equivalent to a linearized
transfer matrix approach\cite{monthus_surface}. If no linearization is performed (this is the so called non-linear transfer approach) than the
method has lead to IDFP behavior for $D \ge 2$, too\cite{monthus_surface}.
Also approximate
renormalization group schemes have been suggested\cite{monthus_garel2,monthus_garel3,monthus_garel4,miyazaki_nishimori},
during which the order of the RG steps is changed in such a way
that the proliferation of new
couplings is avoided. These methods have reproduced some exact 1D results and also provide IDFP behavior
for $D \ge 2$, in agreement with the standard SDRG method.

Most of the results about the critical behavior of the RTIM have been calculated for bulk quantities.
For example the order-parameter of the RTIM is the average magnetization and its value in the bulk, $m_b$,
has the scaling behavior $m_b \sim L^{-x}$, where $L$ is the
linear size of the system and $x$ is the scaling exponent of the bulk magnetization. Real systems, however, have
finite extent and they are limited by boundaries. At a free surface the scaling
behavior of the average surface magnetization, $m_s$, involves a new exponent\cite{binder83,diehl86,pleimling04}, $x_s$.
Due to missing bonds at the surface
there is weaker order, therefore generally $x_s > x$. For the 1D RTIM several properties of the
surface magnetization (the distribution function, average and typical behavior, etc) is exactly known\cite{mccoywu,fisher,bigpaper,monthus_1d}. For example
the surface scaling exponent, $x_s=1/2$, is related to the persistence properties of 1D random walks\cite{bigpaper}.

For higher dimensional RTIM less attention is paid to the calculation of
the surface magnetization: we are aware of one recent work\cite{monthus_surface}, in which the
surface magnetization exponent has been calculated by the non-linear transfer approach. The obtained
values are $x_s=1.2$ and $1.34$, in 2D and 3D, respectively, which are to be compared with the SDRG results
for the bulk magnetization exponent\cite{2dRG,ddRG,ddRG2}: $x=0.98$ and $1.84$, in 2D and 3D, respectively. Since
in 3D the surface magnetization exponent of the non-linear transfer approach is smaller, than the expected
correct value of the bulk magnetization exponent
we conclude that the non-linear transfer approach underestimates the values of $x_s$. Therefore there
is a necessity to obtain more accurate estimates for the surface critical properties of the RTIM.

In this paper we study the boundary critical behavior of the RTIM in higher dimensional systems for
$D=2,~3$ and $4$ by the SDRG method. In the calculation we use the numerical algorithm,
which has been developed in Refs.\cite{ddRG,ddRG2} and has been used to study the bulk critical behavior
of the systems. Besides the surface
magnetization exponent, $x_s$, we calculate local magnetization exponents\cite{ipt93}, which are associated with corners
(in 2D and 3D) as well as with edges (in 3D). We also calculate critical magnetization profiles, when spins are fixed at some
surfaces of the system and study their scaling properties.

The structure of the paper is the following. The essence of the SDRG method and its application to the
calculation of the boundary magnetization is described in Sec.\ref{sec:SDRG}. Our results for 2D, 3D and 4D lattices
are presented in Sec.\ref{sec:results} and discussed in the final section.

\section{SDRG calculation of the local magnetization}
\label{sec:SDRG}
Here we consider the boundary critical properties of the RTIM defined by the Hamiltonian:
\be
{\cal H}=-\sum_{ij} J_{ij} \sigma_i^x \sigma_j^x - \sum_i h_i \sigma_i^z\;
\label{hamiltonian}
\ee
in terms of the $\sigma_i^{x,z}$ Pauli operators at site $i$ of a $D$-dimensional cubic lattice. The $J_{ij}>0$
nearest-neighbor couplings and the $h_i>0$ transverse fields are independent random numbers taken from
the distributions $p(J)$ and $q(h)$, respectively. In this paper we use two different disorder
distributions, which have already been used in Refs.\cite{2dRG,ddRG,ddRG2} for the calculation of the
bulk properties. The advantage of using these distributions is, that the location of the critical points
have already been determined. In the bulk calculations we have observed that the two distributions lead to identical
critical exponents, within the error of the numerical method. Here we assume that universality
holds to local critical properties, too, and check this assumption numerically.
In both type of disorder the couplings are taken from a uniform distribution:
$p(J)=\Theta(J)\Theta(1-J)$, where $\Theta(x)$ is the Heaviside step-function. For the \textit{box-$h$ model}
the transverse fields have a box-like distribution $q(h)=\frac{1}{h_b} \Theta(h)\Theta(h_b-h)$,
whereas for the \textit{fixed-$h$} model the transverse fields are constant: $q(h)=\delta(h-h_f)$.

In the calculations of different local magnetizations (surface, edge and corner) we have used different
finite geometries, in which fixed spin boundary conditions (b.c.) have been used at given planes and
the magnetization profile, $m_l$, is measured perpendicular to the fixed planes. (These are located at $l=1$,
thus $m_1=1$).
For the \textit{surface magnetization} a slab of size $L \times N^{D-1}$ ($L<N$) is used,
and in the short direction we use fixed-free b.c., while
in the other $(D-1)$-directions periodic b.c. is applied. The surface magnetization is given by $m_s=m_{l=L}$,
which scales at the critical point as $m_s \sim L^{-x_s}$.

The \textit{corner magnetization} is measured in 2D at the free corner
of a half square and in 3D at the free corner of a cube, having the shape of a pyramid. (In the
following we use the term pyramid in the 2D case, too.)
The spins at
the base of the pyramid are fixed, while at other surfaces free b.c.-s are used. The magnetization
profile, $m_l$ is measured between the base $(l=1)$ and the corner $(l=L_c=\frac{\sqrt{D}}{2} L)$,
and the corner magnetization is given by $m_c=m_{l=L_c}$. This scales at the critical point as $m_c \sim L^{-x_c}$
with the corner exponent, $x_c$.

\begin{figure}[!ht]
\begin{center}
\includegraphics[width=3.2in,angle=0]{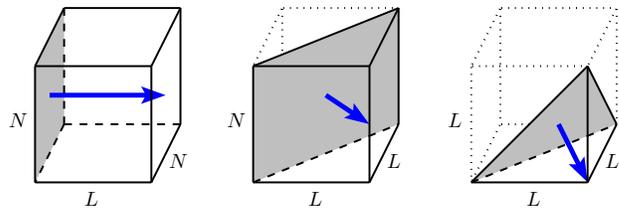}
\end{center}
%\vskip -.5cm
\caption{
\label{fig_1} (Color online) The three geometries used in the calculation of the local
magnetization in 3D. Left panel: slab; middle panel: wedge; right panel: pyramid. Spins at the shaded planes are fixed
and the arrows indicate the directions in which the density profiles are measured.
}
\end{figure}

\textit{Edge magnetization} is calculated in 3D at the free edge of length $N$
of a square column of size $L \times L \times N$, $L<N$. The square column is cut at the square diagonal
plane, thus have the shape of a wedge, and the spins at the base of the wedge are fixed. In the long direction
periodic b.c. is used, while at the other two symmetric surfaces free b.c. is applied. The magnetization
profile, $m_l$ is measured between the base of the wedge ($l=1$) and the free edge ($l=L_e=\frac{\sqrt{2}}{2} L$).
$m_l$ is translationally invariant along the long direction and the edge magnetization is given by $m_e=m_{l=L_e}$.
This scales at the critical point as $m_e \sim L^{-x_e}$, with the edge exponent, $x_e$. The applied
geometries in 3D are shown in Fig. \ref{fig_1}.

%%%%%%%%%% FIG 1  %%%%%%%%%%%%%%%%%%%%%%%%%%%%%%%
%%%%%%%%%% FIG 1  %%%%%%%%%%%%%%%%%%%%%%%%%%%%%%%

We note that the corners and edges we consider here have the specific opening angle: $\pi/2$. The
local critical exponents are generally angle dependent\cite{ipt93}, but we do not study this problem in the present paper.

To calculate the local magnetization in the different geometries
we have used the SDRG method, which is an iterative procedure working in the energy space. At each
step the largest local term of the Hamiltonian, either a coupling, $J_{ij}$, or a transverse field, $h_i$, is decimated
and new terms are generated between the remaining sites in a perturbation calculation. For coupling decimation the
two sites, $i$ and $j$ with original magnetic moments, $\mu_i$ and $\mu_j$, are merged to a new cluster
with an effective moment $\mu'_{ij}=\mu_i+\mu_j$, which is placed in an effective transverse field of strength:
$h'_{ij}=h_ih_j/J_{ij}$. In transverse field decimation the site $i$ is eliminated and its nearest-neighbor sites,
say $j$ and $k$ will be connected by an effective coupling: $J'_{jk}={\rm max}\{J_{ji}J_{ik}/h_i,J_{jk}\}$. In this
last step the so called maximum rule is applied, the use of which is justified at an IDFP.

We apply the numerical algorithm of the SDRG method in Refs.\cite{ddRG,ddRG2}, which has been
used to locate the critical point of the system (for the two forms of the disorder)
and to calculate the bulk critical exponents at the IDFP for different
dimensions, $D=2,~3$ and $4$ (which are disorder independent and listed in Table \ref{table:2}).
To calculate the critical magnetization profile, $m_l$,
we renormalize the system up to the last effective site and
consider the effective cluster, ${\cal C}$, which contains the fixed sites at $l=1$.
If the system has an IDFP, then all spins of ${\cal C}$ are strongly correlated:
in leading order all these spins point to the same direction as at $l=1$,
whereas other sites (not contained in ${\cal C}$) have negligible contribution to the longitudinal magnetization.
Let us denote by $n_l$
the number of sites in ${\cal C}$ at position $l$ and the number of equivalent sites by $\tilde{n}$ (it is
$N^{d-1}$, $N$ and $1$ in the slab, wedge and pyramid geometry, respectively).
The average value of the local magnetization is then given by:
$m_l=\left[ n_l/ \tilde{n}\right]_{\rm av}$, where $[ \dots ]_{\rm av}$ stands for the average over disorder
realizations.

At the critical point the asymptotic form of the magnetization profile is given by scaling considerations. According
to Fisher and de Gennes\cite{fisher78} the decay of the magnetization from the fixed surface is given by:
\be
m_l \sim l^{-x_b},\quad x=x_b,\quad 1 \ll l \ll L\;,
\label{fisher}
\ee
thus it includes the bulk magnetization exponent.
Close to the free endpoint (surface, corner or edge) the magnetization profile has a different power-law decay\cite{binder83}:
\be
m_{l'} \sim (l')^{x_{\alpha b}},\quad 1 \ll l'=L_{\alpha}-l+1 \ll L_{\alpha}\;,
\label{free}
\ee
with $x_{\alpha b}=x_{\alpha}-x$ and $\alpha$ relates to the type of endpoint: $s,~c$ or $e$ (and $L_s \equiv L$).
These relations will be used to obtain independent estimates for the local magnetization exponents. The two scaling
relations in Eqs.(\ref{fisher}) and (\ref{free}) can be incorporated into an interpolation formula:
\be
m_l =\dfrac{A}{L^{x}}\left[\sin(\pi \lambda)\right]^x\left[\cos(\pi \lambda/2)\right]^{x_{\alpha}}\;,
\label{conf}
\ee
with $\lambda=\frac{l}{L_{\alpha}}$. This relation is exact for 1D conformally invariant quantum
systems\cite{burkhardt_xue}. Although the RTIM is not conformally invariant, in 1D Eq.(\ref{conf})
is found to be an excellent approximation\cite{igloi_rieger97,1dprofile}. In the following in the slab geometry ($\alpha=s$)
we shall check the accuracy of Eq.(\ref{conf}) in higher dimensions, too.

\section{Results}
\label{sec:results}
We have calculated the magnetization profiles in the three geometries described in Sec.\ref{sec:SDRG}
in different dimensions: $2 \le D \le 4$ by the SDRG method using two different forms of disorder.
The largest sizes of the systems, the typical
aspect ratios of slabs and wedges as well as the typical number of disorder realizations are collected in Table \ref{table:1}.
Since only a small fraction of samples contains such a correlation cluster, ${\cal C}$, which have also sites at the
free extremity of the system (surface, edge or corner) one should consider a large number of realizations. For
surfaces and edges in a given sample there are several end-point positions, for which
we perform the averages. For corners, however, there is just one end-point in a sample,
therefore one should take even larger number of realizations.
In the following we present our numerical results obtained in different dimensions.

\begin{table}[h!]
\caption{Details of the numerical calculation of the local magnetization. $L_{\text{max}}$:
largest linear size; $N/L$: typical aspect ratio; $N^{\#}$: typical number of realizations. \label{table:1}}
\begin{tabular}{|c|c|c|c|c|c|c|c|c|}  \hline
\multicolumn{1}{|c|}{}&\multicolumn{3}{|c|}{slab} &\multicolumn{2}{|c|}{pyramid}& \multicolumn{3}{|c|}{wedge} \\ \hline
  & $L_{\text{max}}$ & $N/L$ & $N^{\#}$ & $L_{\text{max}}$ & $N^{\#}$ & $L_{\text{max}}$ & $N/L$ & $N^{\#}$ \\ \hline
 2D & 512 & 4. & $10^6$ & 256 & $10^7$ &  &  &   \\ \hline
 3D & 64 &  2. & $10^6$ & 64 & $10^8$ & 64 & 2. & $10^7$  \\ \hline
 4D & 32 &  1.5 & $10^5$ & & &  &  &   \\ \hline
\end{tabular}
\end{table}

\subsection{Calculations in 2D}
\label{sec:2D}
\subsubsection{Surface magnetization}
\label{sec:2D_surf}

%%%%%%%%%% FIG 2  %%%%%%%%%%%%%%%%%%%%%%%%%%%%%%%
\begin{figure}[!ht]
\begin{center}
\includegraphics[width=3.2in,angle=0]{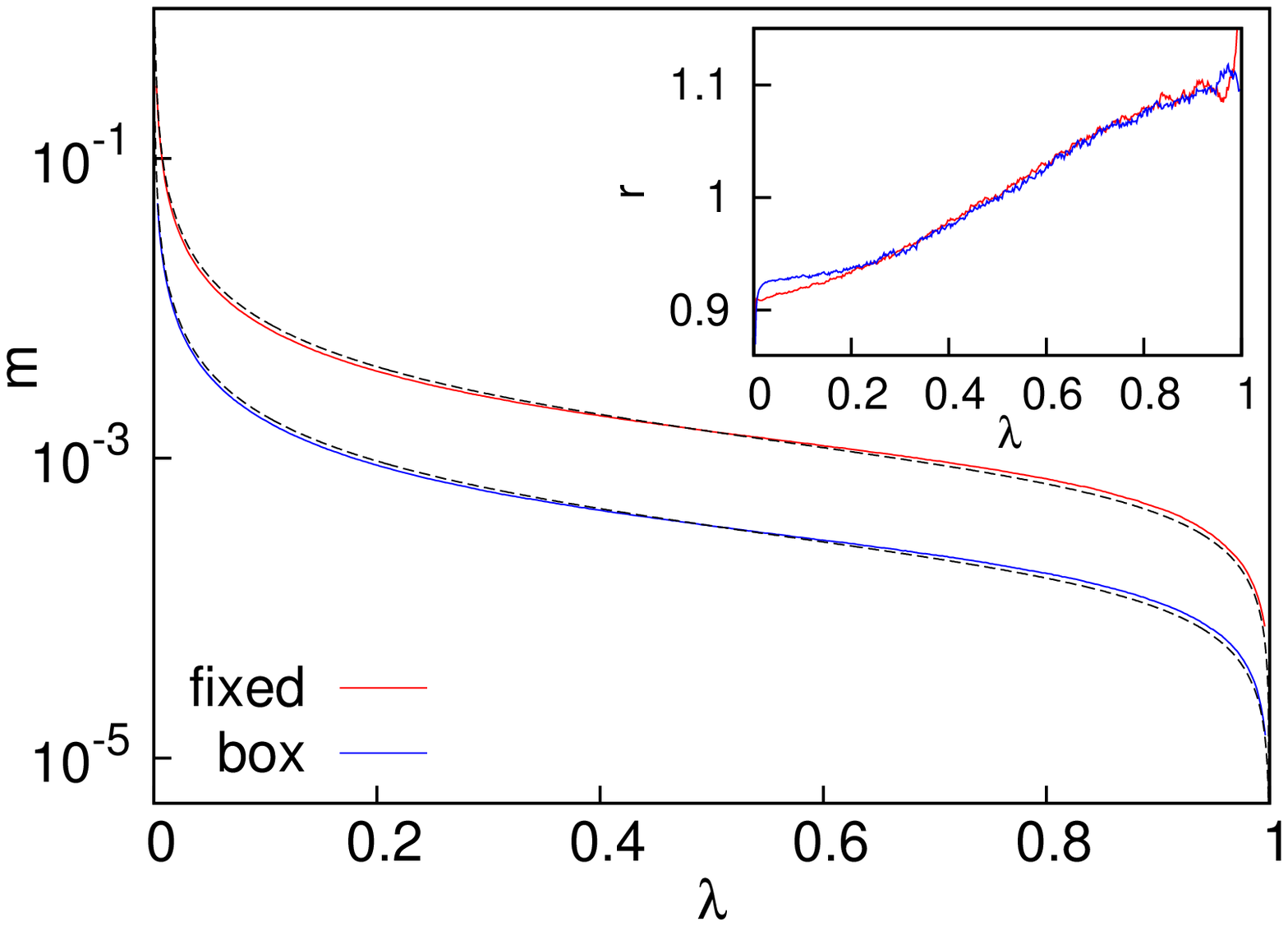}
\end{center}
%\vskip -.5cm
\caption{
\label{fig_2} (Color online) Magnetization profiles in 2D in the slab geometry for fixed-free b.c.-s in
a system of width $L=512$ for box-$h$ and fixed-$h$ randomness. The interpolation formula in Eq.(\ref{conf})
is represented by dashed lines. In the inset the ratio of the magnetization profile and the interpolation formula in Eq.(\ref{conf}) is shown for $x=0.982$, $x_s=1.6$, $A_{\text{fixed}}=1.20$ and $A_{\text{box}}=0.282$.
}
\end{figure}
%%%%%%%%%% FIG 2  %%%%%%%%%%%%%%%%%%%%%%%%%%%%%%%

\begin{figure}[!ht]
\begin{center}
\includegraphics[width=3.2in,angle=0]{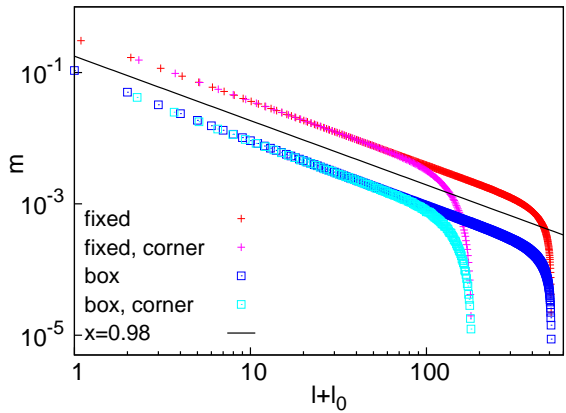}
\end{center}
%\vskip -.5cm
\caption{
\label{fig_3} (Color online) Magnetization profiles near the fixed boundary in 2D for the slab
and the pyramid geometries for the fixes-$h$ and box-$h$ randomness with $L=512$ and $256$, respectively.
In all cases the decay is
characterized by the same exponent, $x_b=0.98(1)$, which according to the Fisher-de Gennes result in Eq.(\ref{fisher}) is
equivalent to the bulk magnetization exponent, see Table \ref{table:2}.}
\end{figure}

\begin{figure}[!ht]
\begin{center}
\includegraphics[width=3.2in,angle=0]{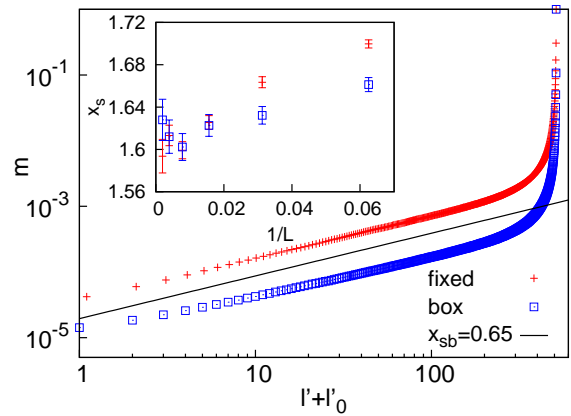}
\end{center}
%\vskip -.5cm
\caption{
\label{fig_4} (Color online) Magnetization profiles near the free boundary in 2D for the slab
geometry for the two type of randomness with $L=512$. In both cases the decay is
characterized by the exponent, $x_{sb}=0.65(2)$. In the inset the finite-size estimates for the
surface magnetization exponent are presented. The extrapolated (disorder independent)
value is given in Table \ref{table:2}.}
\end{figure}

The magnetization profiles in the slab geometry calculated by the two types of disorder are shown in Fig.\ref{fig_2}
as a function of the relative position: $\lambda=l/L$, see Eq.(\ref{conf}).
Here we use a finite-size shift of $l_0={\cal O}(1)$ at the boundaries. 
As already observed in the calculation of the bulk magnetization the typical correlation clusters for fixed-$h$ disorder
contain approximately 6-times more sites, than for box-$h$ disorder. As a consequence the magnetization profiles
are also comparatively larger for fixed-$h$ disorder. As seen in Fig.\ref{fig_2} the magnetization is monotonously
decreasing and the variation is very fast near the two end-points, which are then analyzed in log-log
plots in Figs.\ref{fig_3} and \ref{fig_4}, respectively.

%%%%%%%%%% FIG 3  %%%%%%%%%%%%%%%%%%%%%%%%%%%%%%%
%%%%%%%%%% FIG 3  %%%%%%%%%%%%%%%%%%%%%%%%%%%%%%%

Close to the fixed boundary the magnetization profiles are shown in Fig.\ref{fig_3}, together with
the similar profiles in the pyramid geometry, which will be analyzed in Sec.\ref{sec:2d_corner}.
The magnetization profiles for the two disorder have the same power-law decay and the decay exponent
is estimated from the largest systems as $x_b=0.98(1)$. This is to be compared with the value of the bulk magnetization
exponent $x=0.982(15)$, which has been calculated in Ref.\cite{2dRG} by finite size scaling. We can thus conclude that
the Fisher-de Gennes scaling prediction in Eq.(\ref{fisher}) is well satisfied.

%%%%%%%%%% FIG 4  %%%%%%%%%%%%%%%%%%%%%%%%%%%%%%%
%%%%%%%%%% FIG 4  %%%%%%%%%%%%%%%%%%%%%%%%%%%%%%%

Also at the free-boundary the profiles have a power-law variation (see Fig.\ref{fig_4}) and the corresponding exponent is estimated
from the largest system as: $x_{sb}=0.65(2)$. We have also calculated
the surface magnetization exponent, $x_s$, from the finite-size scaling behavior of the surface magnetization, $m_s$.
Two-point estimates for $x_s$ are presented in the inset of Fig. \ref{fig_4}, which have the same limiting value for large $L$
for the two type of disorder, which is presented in Table \ref{table:2}.
Comparing $x_{sb}$ with $x_s-x$ we can conclude that the scaling prediction in Eq.(\ref{free})
is satisfied.

We have also checked the accuracy of the interpolation formula in Eq.(\ref{conf}) and in the inset of Fig.\ref{fig_2} we have plotted the ratio of the measured profile and the interpolation formula, in which the exponents in Table \ref{table:2}
have been used. As seen in this figure the interpolation formula represents a good approximation, but
the agreement is not perfect, the largest discrepancy is about 10\%.

\subsubsection{Corner magnetization}
\label{sec:2d_corner}
\begin{figure}[!ht]
\begin{center}
\includegraphics[width=3.2in,angle=0]{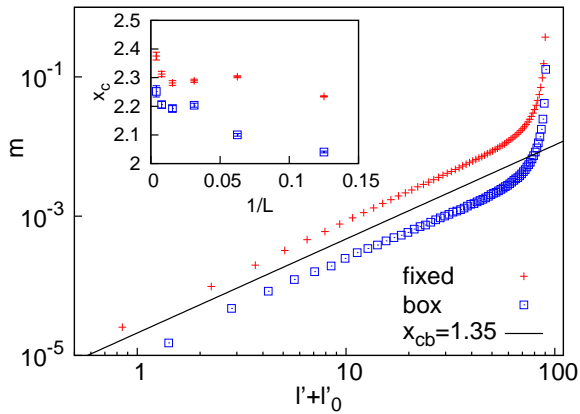}
\end{center}
%\vskip -.5cm
\caption{
\label{fig_5} (Color online) Magnetization profiles in the pyramid geometry near the corner in 2D
for the two type of randomness with $L=256$. In both cases the decay is
characterized by the exponent, $x_{cb}=1.35(10)$. In the inset the finite-size estimates for the
corner magnetization exponent are presented\cite{note}. The extrapolated (disorder independent)
value is given in Table \ref{table:2}.}
\end{figure}

The calculations are performed in the pyramid geometry and the critical magnetization profile close to the
fixed plane is shown in Fig.\ref{fig_2} in a log-log plot for the two different initial disorder. As
discussed in Sec.\ref{sec:2D_surf} in
this figure also the profiles in the slab geometry are presented and the two types of profiles
are very close to each other: they are indistinguishable within the error of the calculation.
Thus in agreement with scaling theory the decay of the profile in
the pyramid geometry is in a power-law form with a decay exponent, $x_b=x$. The magnetization profile
at the other end, i.e. starting from the corner is shown in Fig.\ref{fig_5} and the
corresponding decay exponent of the magnetization, $x_{cb}$, is presented in Table \ref{table:2}.

%%%%%%%%%% FIG 5  %%%%%%%%%%%%%%%%%%%%%%%%%%%%%%%
%%%%%%%%%% FIG 5  %%%%%%%%%%%%%%%%%%%%%%%%%%%%%%%

From finite-size scaling the corner magnetization exponent, $x_c$, is calculated by two-point fit and
the effective, size-dependent exponents are presented in the inset of Fig.\ref{fig_5} for the two different type of disorder.
The extrapolated value which is disorder independent is given in Table \ref{table:2}.

\subsection{Calculations in 3D}
\label{sec:3D}
\subsubsection{Surface magnetization}
\label{sec:3D_surf}

%%%%%%%%%% FIG 6  %%%%%%%%%%%%%%%%%%%%%%%%%%%%%%%
\begin{figure}[!ht]
\begin{center}
\includegraphics[width=3.2in,angle=0]{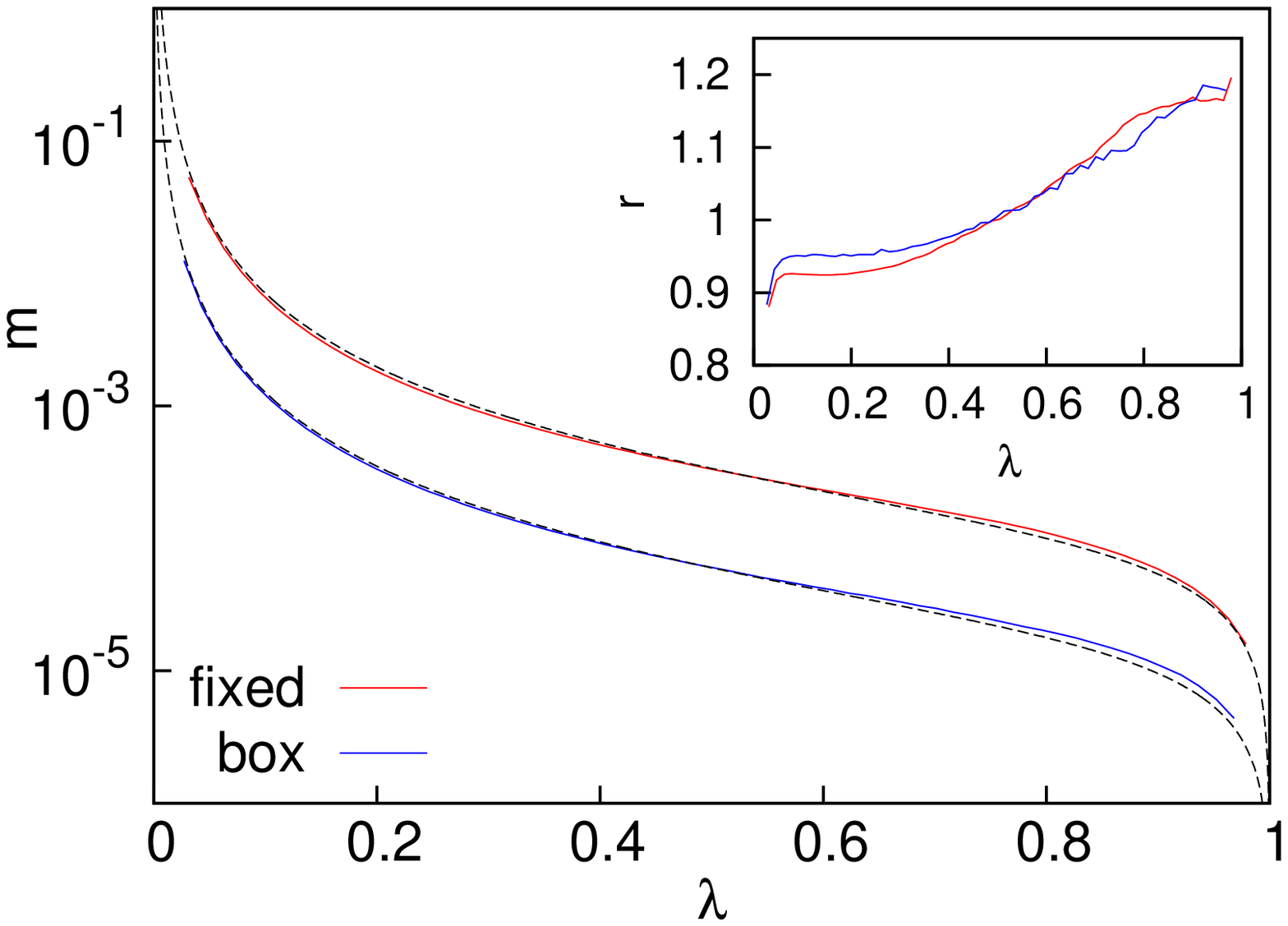}
\end{center}
%\vskip -.5cm
\caption{
\label{fig_6} (Color online) Magnetization profiles in 3D in the slab geometry for fixed-free b.c.-s in
a system of width $L=64$ for box-$h$ and fixed-$h$ randomness. The interpolation formula in Eq.(\ref{conf})
is represented by dashed lines. In the inset the ratio of the magnetization profile and the interpolation formula in Eq.(\ref{conf}) is shown for $x=1.84$, $x_s=2.65$, $A_{\text{fixed}}=1.78$ and $A_{\text{box}}=0.316$.}
\end{figure}
%%%%%%%%%% FIG 6  %%%%%%%%%%%%%%%%%%%%%%%%%%%%%%%

\begin{figure}[!ht]
\begin{center}
\includegraphics[width=3.2in,angle=0]{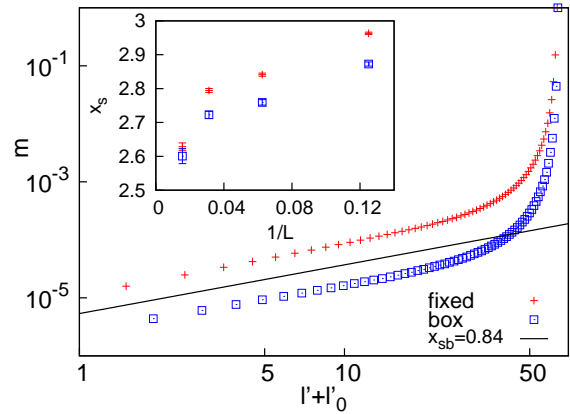}
\end{center}
%\vskip -.5cm
\caption{
\label{fig_7} (Color online) Magnetization profiles near the free boundary in 3D for the slab
geometry for the two types of randomness with $L=64$. In both cases the decay is
characterized by the exponent, $x_{sb}=0.84(7)$. In the inset the finite-size estimates for the
surface magnetization exponent are presented. The extrapolated (disorder independent)
value is given in Table \ref{table:2}.}
\end{figure}

The magnetization profiles in the slab geometry are shown in Fig. \ref{fig_6} for the two types of disorder.
Close to the fixed boundary the exponent associated to the decay of the profile is estimated as $x_b=1.855(20)$ which
agrees with the finite-size estimate of the bulk magnetization exponent, see Table \ref{table:2}. Near the free
surface the profiles for the two types of disorder are shown in Fig. \ref{fig_7} and the estimated decay
exponent, $x_{sb}$, is presented in Table \ref{table:2}.

%%%%%%%%%% FIG 7  %%%%%%%%%%%%%%%%%%%%%%%%%%%%%%%
%%%%%%%%%% FIG 7  %%%%%%%%%%%%%%%%%%%%%%%%%%%%%%%

The surface magnetization exponent is estimated through finite-size scaling and the effective, size-dependent
values are shown in the inset of Fig. \ref{fig_7} for the two types of disorder. The extrapolated exponent is
disorder independent and given in Table \ref{table:2}. We conclude that the scaling relation in Eq.(\ref{free}) is satisfied
within the error of the calculation.

We have checked the accuracy of the interpolation formula in Eq.(\ref{conf}) and the ratio of the measured
profile and the interpolation formula is shown in the inset of Fig. \ref{fig_6}. Also in this case Eq.(\ref{conf})
is a good approximation, the maximal discrepancy is somewhat larger, than in the 2D case, see in Fig. \ref{fig_2}.

\subsubsection{Edge magnetization}
\label{sec:3d_edge}

\begin{figure}[!ht]
\begin{center}
\includegraphics[width=3.2in,angle=0]{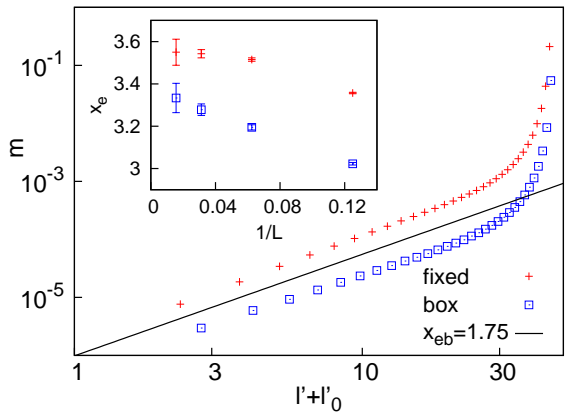}
\end{center}
%\vskip -.5cm
\caption{
\label{fig_8} (Color online) Magnetization profiles near the free edge in 3D in the wedge
geometry for the two type of randomness with $L=64$. In both cases the decay is
characterized by the exponent, $x_{eb}=1.75(15)$. In the inset the finite-size estimates for the
edge magnetization exponent are presented\cite{note}. The extrapolated (disorder independent)
value is given in Table \ref{table:2}.}
\end{figure}

We have measured the magnetization profile in the wedge geometry and here we analyze its behavior
close to the free edge, see Fig. \ref{fig_8}.

%%%%%%%%%% FIG 8  %%%%%%%%%%%%%%%%%%%%%%%%%%%%%%%
%%%%%%%%%% FIG 8  %%%%%%%%%%%%%%%%%%%%%%%%%%%%%%%

The estimated decay exponent, $x_{eb}$, is presented in Table \ref{table:2} together with the
extrapolated value of the edge exponent, $x_e$, for which the finite-size estimates are
shown in the inset of Fig. \ref{fig_8}. In this case, too the scaling relation in Eq.(\ref{free}) is satisfied.

\subsubsection{Corner magnetization}
\label{sec:3d_corner}

%%%%%%%%%% FIG 9  %%%%%%%%%%%%%%%%%%%%%%%%%%%%%%%
\begin{figure}[!ht]
\begin{center}
\includegraphics[width=3.2in,angle=0]{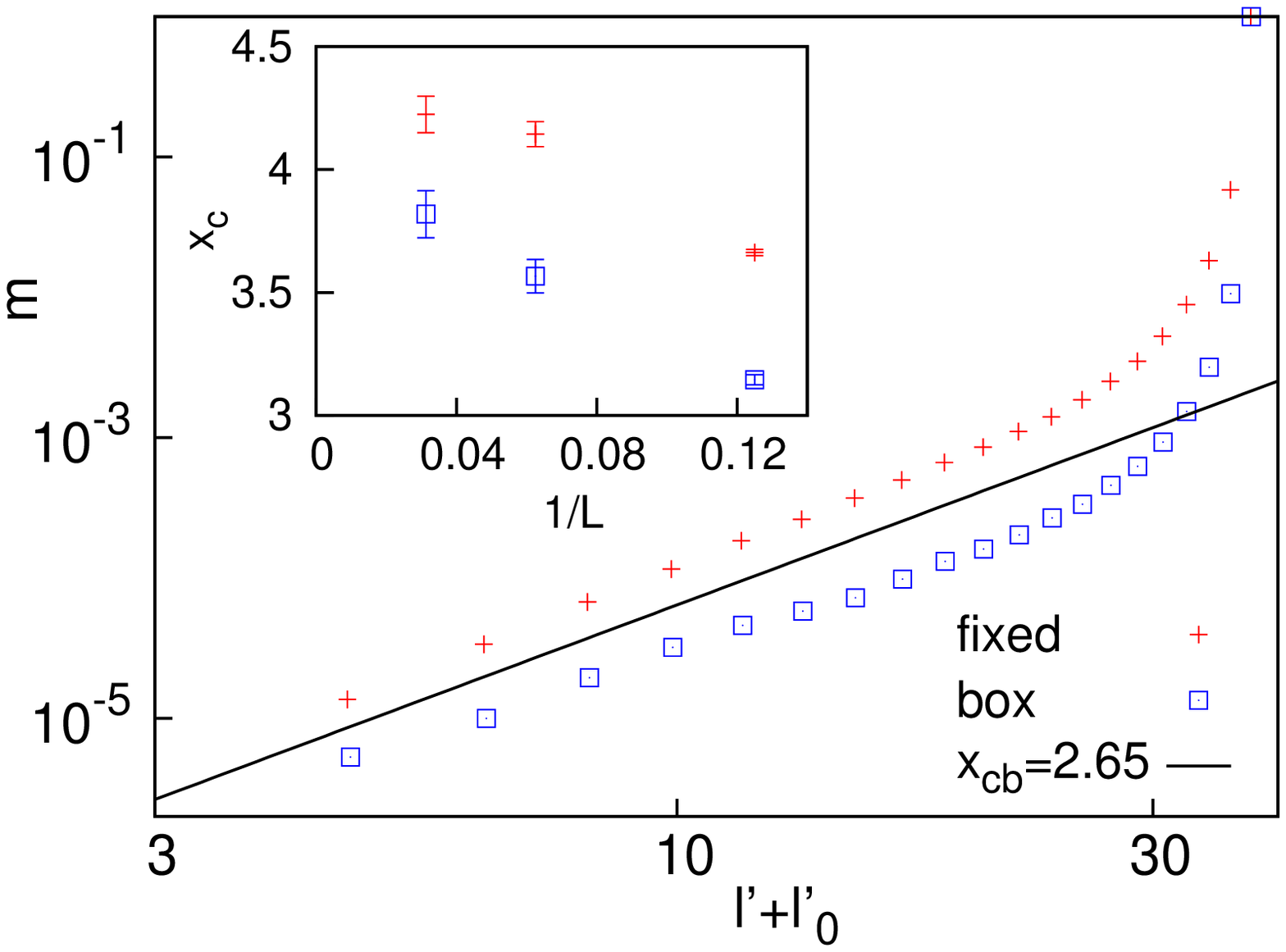}
\end{center}
%\vskip -.5cm
\caption{
\label{fig_9} (Color online) Magnetization profiles near the free corner in 3D in the pyramid
geometry for the two type of randomness with $L=64$. In both cases the decay is
characterized by the exponent, $x_{cb}=2.65(25)$. In the inset the finite-size estimates for the
edge magnetization exponent are presented up to $L=32$\cite{note}. The extrapolated (disorder independent)
value is given in Table \ref{table:2}.}
\end{figure}
%%%%%%%%%% FIG 9  %%%%%%%%%%%%%%%%%%%%%%%%%%%%%%%

We close our study in 3D by calculating the magnetization profile in the pyramid geometry: the result is
shown in Fig. \ref{fig_9} close to the free corner. (In the inset finite-size estimates of the corner exponent
are presented.)
Estimates of the decay exponent, $x_{cb}$, and the corner exponent, $x_c$, are presented
in Table \ref{table:2}, which satisfy the scaling relation in Eq.(\ref{free}).

\subsection{Calculations in 4D}
\label{sec:4D}

\begin{figure}[!ht]
\begin{center}
\includegraphics[width=3.2in,angle=0]{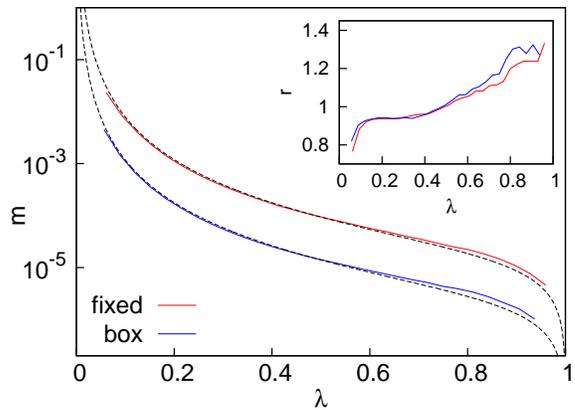}
\end{center}
%\vskip -.5cm
\caption{
\label{fig_10} (Color online) Magnetization profiles in 4D in the slab geometry for fixed-free b.c.-s in
a system of width $L=32$ for box-$h$ and fixed-$h$ randomness. The interpolation formula in Eq.(\ref{conf})
is represented by dashed lines.
In the inset the ratio of the magnetization profile and the interpolation formula is shown for $x=2.72$, $x_s=3.7$, $A_{\text{fixed}}=4.19$ and $A_{\text{box}}=0.625$ . %In the interpolation formula we used the shifts: $l_0^{free}=$ and $l_0^{fix}=$.
}
\end{figure}

In 4D the available system sizes are limited, see Table \ref{table:1}, therefore we could only
study the magnetization profile in the slab geometry, which is shown in Fig. \ref{fig_10}.

%%%%%%%%%% FIG 10  %%%%%%%%%%%%%%%%%%%%%%%%%%%%%%%
%%%%%%%%%% FIG 10  %%%%%%%%%%%%%%%%%%%%%%%%%%%%%%%

\begin{figure}[!ht]
\begin{center}
\includegraphics[width=3.2in,angle=0]{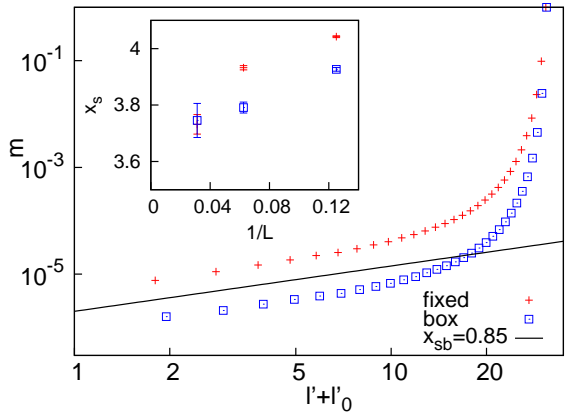}
\end{center}
%\vskip -.5cm
\caption{
\label{fig_11} (Color online) Magnetization profiles near the free boundary in 4D for the slab
geometry for the two type of randomness with $L=32$. In both cases the decay is
characterized by the exponent, $x_{sb}=0.85(15)$. In the inset the finite-size estimates for the
surface magnetization exponent are presented. The extrapolated (disorder independent)
value is given in Table \ref{table:2}.}
\end{figure}

Close to the fixed surface the decay
exponent is calculated as $x_b=2.72(10)$, which agrees well with the finite-size estimate
of the bulk magnetization exponent, see Table \ref{table:2}.
The magnetization profile close to the free surface is shown in Fig. \ref{fig_11}.

%%%%%%%%%% FIG 11  %%%%%%%%%%%%%%%%%%%%%%%%%%%%%%%
%%%%%%%%%% FIG 11  %%%%%%%%%%%%%%%%%%%%%%%%%%%%%%%

Estimates of the decay exponent, $x_{sb}$, and that of the surface magnetization exponent, $x_s$,
which are presented in Table \ref{table:2} contain somewhat larger errors, than in lower dimensional
calculations. However the scaling relation in Eq.(\ref{free}) is satisfied in this case, too.
Also the interpolation formula in Eq.(\ref{conf}) is a good approximation as can be seen in the inset
of Fig. \ref{fig_10}.

\section{Discussion}
\label{sec:disc}
In this paper we have used the SDRG method to calculate the magnetization profiles of the random transverse Ising model
in 2D, 3D and 4D in different geometries: slab, corner and wedge having a fixed surface.
At the critical point decay exponents are calculated both at the fixed end and at the free end of the
profiles. These exponents, which are presented in Table \ref{table:2} are found to be disorder
independent, at least for strong enough disorder, for which the critical properties of the system
are controlled by an IDFP. From finite-size scaling studies of the local magnetization at the free ends of the
profile local (surface, corner and edge) critical exponents are calculated, see Table \ref{table:2}.
For all types of profiles considered here the scaling relations in Eqs.(\ref{fisher}) and (\ref{free})
are satisfied and the interpolation formula in Eq.(\ref{conf}) is found to be a good approximation in
the slab geometry.
Our results, concerning the properties of the average local magnetization of the RTIM are rather complete, these are
comparable with the existing results in the non-random system.

By the SDRG method the average local magnetization, $m_l$, is obtained as the ratio of such rare realizations, in which the
correlation cluster contains the given site. By this method the typical value of the local magnetization
could be estimated by the strength of the effective coupling, $J'_{1l}$, which is generated
between the fixed surface and the site. For surface spins it scales as $m_s^{typ} \sim J'_{1L} \sim \exp(-A L^{\psi})$,
where $\psi$ is a characteristic exponent in the IDPF\cite{im,danielreview}, which has been calculated in Refs.\cite{2dRG,ddRG,ddRG2}. Thus by calculating $m_s^{typ}$ by some other means one can obtain independent estimates for the exponent $\psi$.

Concerning the dimensional dependence of the average surface magnetization exponent, we write it in the form:
$x_s(D)=D_s+p(D)$, where $D_s=D-1$ is the dimension of the surface and $p(D)$ is a number close to $1/2$.
In 1D $p(1)=1/2$ is shown to be the persistence exponent of the random walk\cite{bigpaper} and we propose here an analogous explanation for $D>1$, too. Let us denote by $\mu_s$ the number of surface points of the correlation cluster, ${\cal C}$ which starts at the fixed boundary. We have checked, that $\mu_s$ has an exponential distribution: $P(\mu_s) \sim \exp(-\mu_s/\tilde{\mu})$,
with $\tilde{\mu} \approx B (N/L)^{D-1}$, c.f. in 2D we have $B=0.5$ and $B=1.0$, for fixed-$h$ and box-$h$
disorder, respectively. Consequently the surface points of ${\cal C}$ are grown from uncorrelated domains.
The average number of surface points in an area $L^{D-1}$ then scales as $[\mu_s]_{\rm av} \sim L^{-p(D)}$.
As explained in Ref.\cite{ddRG2} (see Fig. 5 there)
the correlation cluster is embedded into a connected subgraph, which contains all the decimated
sites (the results of both $h$- and $J$-decimations) and which is related to a low-energy excitation
of the system. The number of points in the connected subgraph is $\tilde{L} \sim L^{D_f}$,
where the fractal dimension, $D_f$, is close to one. If we replace
the connected subgraph with a linear chain with $\tilde{L}$ sites then we obtain from the random walk result:
$p(D) \approx D_f/2$. Indeed our numerical estimates\cite{phd} of $D_f$ and the surface magnetization scaling dimensions
in Table \ref{table:2} are in agreement with this relation. In higher dimensions we
expect that the structure of the SDRG transformation, in particular the topology of the connected clusters
follows the trend observed in this paper, thus the surface magnetization exponent generally obeys the relation:
$x_s(D+1)-x_s(D) \approx 1$. Based on this result we expect that a
simplified SDRG procedure, such as a modified version of those used in Refs.\cite{dimitrova_mezard,monthus_garel1,monthus_surface,monthus_garel2,monthus_garel3,monthus_garel4,miyazaki_nishimori}
can be constructed, which captures
the main results about the surface critical properties of the RTIM.

\begin{widetext}
\begin{table}[h]
\caption{Estimates of the critical exponents obtained by finite-size scaling: $x$, $x_s$, $x_c$, $x_e$
and the exponents associated with the decay of the profile: $x_b$, $x_{sb}$, $x_{cb}$, $x_{eb}$. ($x$ is
taken from Refs.\cite{2dRG,ddRG,ddRG2} and the exact results in 1D are from Refs.\cite{fisher,mccoywu,bigpaper}).
For the pure system (see c.f. in\cite{pleimling04}) $x_s=0.5$ and $1.27$ for 1D and 2D, respectively and for
$D\ge 3$ the mean-field
result holds: $x_s=(D+1)/2$. The corner exponent in 2D is $x_c=2.06$.
\label{table:2}}
\begin{tabular}{|c||c|c||c|c||c|c||c|c|}  \hline
\multicolumn{1}{|c||}{}&\multicolumn{2}{|c||}{bulk} &\multicolumn{2}{|c||}{surface} &\multicolumn{2}{|c||}{corner}& \multicolumn{2}{|c|}{edge} \\ \hline
  & $x$ & $x_b$ & $x_s$ & $x_{sb}$ & $x_c$ & $x_{cb}$ & $x_e$ & $x_{eb}$ \\ \hline
 1D & $(3-\sqrt{5})/4$ &  &  $0.5$ &  & &  &  &   \\ \hline
 2D &0.982(15) & 0.98(1)  &1.60(2) & 0.65(2) & 2.3(1) & 1.35(10) &  &   \\ \hline
 3D & 1.840(15) &1.855(20)  & 2.65(15) & 0.84(7) & 4.2(2) & 2.65(25) & 3.50(15) & 1.75(15)  \\ \hline
 4D & 2.72(12) & 2.72(10) &  3.7(1) & 0.85(15) & &  &  &   \\ \hline
\end{tabular}
\end{table}
\end{widetext}

Our results about the local critical behavior of the RTIM are relevant to other random quantum magnets having
discrete symmetry, we mention the random quantum Potts\cite{senthil}, clock and Ashkin-Teller models\cite{carlon-clock-at}.
Also the surface, corner and/or edge
exponents of the random contact process are expected to be given by the RTIM values in Table \ref{table:2}. To check
this conjecture one should repeat recent Monte Carlo simulations about this model\cite{vojta09,vojta12}.

Our studies of the local critical behavior can be extended in different directions. For example, one can measure
the corner and edge magnetization exponents at different opening angles or one can consider anisotropic systems,
in which the distribution of disorder is different in the different directions. One can also consider the surface
critical behavior in the presence of enhanced surface couplings, in which case the so called extraordinary
and surface transitions\cite{binder83,diehl86,pleimling04} could be studied, too. 

\begin{acknowledgments}
This work has been supported by the Hungarian National Research Fund under grant
No OTKA K75324 and K77629.
\end{acknowledgments}


\begin{thebibliography}{99}
\vskip -.5cm
\bibitem{im} 
For a review, see: F. Igl\'oi and C. Monthus, Physics Reports {\bf 412}, 277, (2005).

\bibitem{mdh}
        S.K. Ma, C. Dasgupta and C.-K. Hu, Phys. Rev. Lett. {\bf 43}, 1434 (1979);
        C. Dasgupta and S.K. Ma, Phys. Rev. B{\bf 22}, 1305 (1980).

\bibitem{fisher}
        D.S. Fisher, Phys. Rev. Lett. {\bf 69}, 534 (1992); 
        Phys. Rev. B {\bf 51}, 6411 (1995).

\bibitem{danielreview}
D.S. Fisher, Physica A {\bf 263}, 222 (1999)

\bibitem{mccoywu}
       B. M. McCoy and T. T. Wu,
       Phys. Rev. \textbf{176}, 631 (1968);
       Phys. Rev. \textbf{188}, 982 (1969);
       B. M. McCoy,
       Phys. Rev. \textbf{188}, 1014 (1969);
       Phys. Rev. B \textbf{2}, 2795 (1970).


\bibitem{shankar}
       R. Shankar and G. Murthy,
       Phys. Rev. B \textbf{36}, 536 (1987).

\bibitem{young_rieger96}
A. P. Young and H. Rieger, Phys. Rev. B \textbf{53}, 8486 (1996).

\bibitem{igloi_rieger97}
        F. Igl\'oi and H. Rieger, Phys. Rev. Lett. {\bf 78}, 2473 (1997).

\bibitem{bigpaper}
        F. Igl\'oi and H. Rieger, Phys. Rev. B{\bf 57} 11404 (1998).

\bibitem{motrunich00}
O. Motrunich, S.-C. Mau, D.A. Huse and D.S. Fisher, \prb \textbf{61}, 1160 (2000).

\bibitem{lin00} Y.-C. Lin, N. Kawashima, F. Igl\'oi and H. Rieger, Progress in Theor. Phys. {\bf 138}, (Suppl.) 479 (2000).

\bibitem{karevski01} D. Karevski, Y-C. Lin, H. Rieger, N. Kawashima and F. Igl\'oi, Eur. Phys. J. B \textbf{20} 267 (2001).

\bibitem{lin07} Y-C.Lin, F. Igl\'oi and H. Rieger, \prl \textbf{99}, 147202 (2007).

\bibitem{yu07} R. Yu, H. Saleur and S. Haas, \prb \textbf{77}, 140402 (2008).

\bibitem{ladder} I. A. Kov\'acs and F. Igl\'oi, Phys. Rev. B \textbf{80}, 214416 (2009).

\bibitem{2dRG} I. A. Kov\'acs and F. Igl\'oi, Phys. Rev. B \textbf{82}, 054437 (2010).

\bibitem{ddRG}  I. A. Kov\'acs and F. Igl\'oi, Phys. Rev. B \textbf{83}, 174207 (2011).

\bibitem{ddRG2}  I. A. Kov\'acs and F. Igl\'oi, J. Phys. Condens. Matter \textbf{23}, 404204 (2011).

\bibitem{phd} I. A. Kov\'acs, PhD thesis (2012).

\bibitem{kovacs_igloi12} I. A. Kov\'acs and F. Igl\'oi, EPL \textbf{97}, 67009 (2012).

\bibitem{pich98} C. Pich, A.P. Young, H. Rieger and N. Kawashima, \prl \textbf{81}, 5916 (1998).

\bibitem{vojta09} T. Vojta, A. Farquhar and J. Mast, \pre \textbf{79}, 011111 (2009).

\bibitem{vojta12} T. Vojta, arXiv:1209.1400 (2012).

\bibitem{hiv} J. Hooyberghs, F. Igl\'oi and C. Vanderzande, Phys. Rev. Lett. {\bf 90} 100601, (2003); Phys. Rev. E {\bf 69}, 066140 (2004).

\bibitem{dimitrova_mezard}
O. Dimitrova and M. M\'ezard, J. Stat. Mech. P01020 (2011).

\bibitem{monthus_garel1} C. Monthus and Th. Garel, J. Phys. A: Math. Theor. \textbf{45}, 095002 (2012).

\bibitem{monthus_surface} C. Monthus and Th. Garel, J. Stat. Mech. P01008 (2012).

\bibitem{monthus_garel2} C. Monthus and Th. Garel, J. Stat. Mech. P05002 (2012).

\bibitem{monthus_garel3} C. Monthus and Th. Garel, J. Stat. Mech. P10010 (2012).

\bibitem{monthus_garel4} C. Monthus and Th. Garel, J. Stat. Mech. P09016 (2012).

\bibitem{miyazaki_nishimori} R. Miyazaki and H. Nishimori, arXiv:1210.5053 (2012).

\bibitem{cavity}
L. B. Ioffe and M. M\'ezard, Phys. Rev. Lett. \textbf{105}, 037001 (2010);
F. M. Feigel'man, L. B. Ioffe and M. M\'ezard, Phys. Rev. B \textbf{82}, 184534 (2010).

\bibitem{binder83} K. Binder, in \textit{Phase Transitions and Critical Phenomena}, edited by C. Domb and J.~L. Lebowitz (Academic, London, 1983), Vol.~8, p.~1.

\bibitem{diehl86} H.~W. Diehl in \textit{Phase Transitions and Critical Phenomena}, edited by C. Domb and
J.~L. Lebowitz (Academic Press, London, 1986), Vol.~10, p.~75.

\bibitem{pleimling04} M. Pleimling, J. Phys. A {\bf 37}, R79 (2004).

\bibitem{monthus_1d} C. Monthus, Phys. Rev. B \textbf{69}, 054431 (2004)

\bibitem{ipt93} F. Igl\'oi, I. Peschel, and L. Turban, Advances in Physics {\bf 42}, 683 (1993).

\bibitem{fisher78} M.E. Fisher, and P-G. de Gennes, C. R. Acad. Sci. (Paris) {\bf 287}, 207 (1978).

\bibitem{burkhardt_xue} T. W. Burkhardt and T. Xue, Phys. Rev. Lett. \textbf{66}, 895 (1991).

\bibitem{1dprofile} M. Karsai, I. A. Kov\'acs, J-Ch. Angles d'Auriac and F. Igl\'oi Phys. Rev. E \textbf{78}, 061109 (2008).

\bibitem{note} For the largest size the discrepancy between the estimates for the two types of disorder
is presumably due to the limited accuracy of the position of the critical points,
what can be obtained with our numerical method.

\bibitem{senthil}
       T. Senthil and S. N. Majumdar
       Phys. Rev. Lett. \textbf{76}, 3001 (1996)

\bibitem{carlon-clock-at} E. Carlon, P. Lajk\'o, and F. Igl\'oi, Phys. Rev.
  Lett. \textbf{87}, 277201 (2001)

\end{thebibliography}
\end{document}